\def\paperauthorA{Philippe Esling}
\def\paperauthorB{Axel Chemla--Romeu-Santos}
\def\paperauthorC{Adrien Bitton}

\documentclass[twoside,a4paper]{article}
\usepackage{dafx_18}
\usepackage{amsmath,amssymb,amsfonts,amsthm}
\usepackage{euscript}
\usepackage[latin1]{inputenc}
\usepackage[T1]{fontenc}
\usepackage{ifpdf}
\usepackage[ruled,linesnumbered]{algorithm2e}
\usepackage{multirow}
\usepackage{makecell}
\usepackage{hyperref}

\usepackage[english]{babel}
\usepackage{caption}
\usepackage{subfig} 
\usepackage{color}

\ninept

\usepackage{times}

\newif\ifpdf
\ifx\pdfoutput\relax
\else
   \ifcase\pdfoutput
      \pdffalse
   \else
      \pdftrue
\fi

\ifpdf 
  \usepackage[pdftex]{graphicx}
  \usepackage[figure,table]{hypcap}
\else 
  \usepackage[dvips]{epsfig,graphicx}
  \usepackage[figure,table]{hypcap}
\fi

\newcommand\given[1][]{\:#1\vert\:}

\DeclareMathOperator*{\argmin}{arg\,min}
\newcommand{\norm}[1]{\left\lVert#1\right\rVert}

\title{Generative timbre spaces: regularizing variational auto-encoders with perceptual metrics}

\affiliation{
\paperauthorA\textsuperscript{1}\sthanks{This work was supported by project MAKIMOno 17-CE38-0015-01 funded by the French ANR and Canadian NSERC (STPG 507004-17) and the ACTOR Partnership funded by the Canadian SSHRC (895-2018-1023).}\, \paperauthorB\textsuperscript{2,1}, \paperauthorC\textsuperscript{1}}
{
\textsuperscript{1}\href{http://www.ircam.fr/}{IRCAM}, UMR 9912 CNRS - Sorbonne Universite, Paris, France\\ 
\textsuperscript{2}Laboratorio d'Informatica Musicale - Universita degli Studi di Milano, Italy\\
{\tt \href{mailto:esling@ircam.fr}{esling@ircam.fr}}
}

\begin{document}

\maketitle

\begin{abstract}
Timbre spaces have been used in music perception to study the perceptual relationships between instruments based on dissimilarity ratings. However, these spaces do not generalize to novel examples and do not provide an invertible mapping, preventing audio synthesis. In parallel, generative models have aimed to provide methods for synthesizing novel timbres. However, these systems do not provide an understanding of their inner workings and are usually not related to any perceptually relevant information.\\
Here, we show that Variational Auto-Encoders (VAE) can alleviate all of these limitations by constructing \emph{generative timbre spaces}. To do so, we adapt VAEs to learn an audio latent space, while using perceptual ratings from timbre studies to regularize the organization of this space. The resulting space allows us to analyze novel instruments, while being able to synthesize audio from any point of this space. We introduce a specific regularization allowing to enforce any given similarity distances onto these spaces. We show that the resulting space provide almost similar distance relationships as timbre spaces. We evaluate several spectral transforms and show that the Non-Stationary Gabor Transform (NSGT) provides the highest correlation to timbre spaces and the best quality of synthesis. Furthermore, we show that these spaces can generalize to novel instruments and can generate any path between instruments to understand their timbre relationships. As these spaces are continuous, we study how audio descriptors behave along the latent dimensions. We show that even though descriptors have an overall non-linear topology, they follow a locally smooth evolution. Based on this, we introduce a method for \emph{descriptor-based synthesis} and show that we can control the descriptors of an instrument while keeping its timbre structure.
\end{abstract}
\section{Introduction}

For the past decades, music perception research has tried to understand the perception of instrumental \emph{timbre}.  Timbre is the set of properties that distinguishes two instruments that play the same note at the same intensity. To do so, several studies \cite{mcadams2006meta} collected human dissimilarity ratings between pairs of audio samples inside a set of instruments. These ratings are organized by applying MultiDimensional Scaling (MDS), leading to \emph{timbre spaces}, which exhibit the perceptual similarities between different instruments. By analyzing the dimensions of resulting spaces, the studies tried to correlate audio descriptors to the perception of timbre \cite{grey1978perceptual}.  Although these spaces provided interesting avenues of analysis, they are inherently limited by the fact that ordination techniques (e.g. MDS) produce a fixed space, which has to be recomputed entirely for any new sample. Therefore, these spaces do not generalize to novel examples and do not provide an invertible mapping, precluding audio synthesis to understand their perceptual topology.

In parallel, recent developments in audio synthesis using \emph{generative models} has seen great improvements with the introduction of approaches such as the \emph{WaveNet} \cite{van2016wavenet} and \emph{SampleRNN} \cite{mehri2016samplernn} architectures. These allow to generate novel high-quality audio matching the properties of the corpus they have been trained on. However, these models give little cue and control over the output or the features it results from. More recently, \emph{NSynth} \cite{engel2017neural} has been proposed to synthesize audio by allowing to morph between specific instruments. However, these models still require very large number of parameters, long training times and a large number of examples. Amongst recent generative models, another key proposal is the \textit{Variational Auto-Encoder} (\textit{VAE}) \cite{kingma2013auto}. In these, a \emph{latent space} is learned that allows both to encode data for analysis, but also to sample from it in order to generate novel content. VAEs address the limitations of control and analysis through this latent space, while remaining simple and fast to learn with a small set of examples. Furthermore, VAEs seem able to disentangle underlying variation factors by learning independent latent variables accounting for distinct generative processes \cite{higgins2016beta}. However, these latent dimensions are learned in an unsupervised way. Therefore, they are not related to perceptual properties, which might hamper their understandability or their use for audio analysis and synthesis.

Here, we show that we can bridge timbre perception analysis and perceptually-relevant audio synthesis by regularizing the learning of VAE latent spaces so that they match the perceptual distances collected from timbre studies. Our overall approach is depicted in Figure~1. First, we adapt the VAE to analyze musical audio content, by comparing the use of different spectral transforms as input to the learning. We show that, amongst the Short-Term Fourier Transform (STFT), Discrete Cosine Transform (DCT) and the Non-Stationary Gabor Transform (NSGT) \cite{balazs2011theory}, the NSGT provides the best reconstruction abilities and regularization performances. By training this model on a small database of spectral frames, it already provides a generative model with an interesting latent space, able to synthesize novel instrumental timbres.
Then, we introduce a regularization to the learning objective inspired by the t-Stochastic Neighbors Embedding (t-SNE) \cite{maaten2008visualizing}, aiming to enforce that the latent space exhibits the same distances between instruments as those found in timbre studies. To do so, we build a model of perceptual relationships by analyzing dissimilarity ratings from five independent timbre studies \cite{grey1977multidimensional, krumhansl1989musical, iverson1993isolating, mcadams1995perceptual, lakatos2000common}. We show that perceptually-regularized latent spaces are simultaneously coherent with perceptual ratings, while being able to synthesize high-quality audio distributions. Hence, we drive the learning of latent spaces to match the topology of given target spaces. 

We demonstrate that these spaces can be used for generating novel audio content, by analyzing their reconstruction quality on a test dataset. Furthermore, we show that paths in the latent space (where each point corresponds to a single spectral frame) provide sound synthesis with continuous evolutions of timbre. We also show that these spaces generalize to novel samples, by encoding a set of instruments that were not part of the training set. Therefore, the spaces could be used to predict the perceptual similarities of novel instruments. Finally, we study how traditional audio descriptors are organized along the latent dimensions. We show that even though descriptors behave in a non-linear way across space, they still follow a locally smooth evolution. Based on this smoothness property, we introduce a method for \emph{descriptor-based path synthesis}. We show that we can modify an instrumental distribution so that it matches a given target evolution of audio descriptors, while remaining perceptually smooth. The source code, audio examples and  animations are available on a supporting repository\footnote{\url{https://github.com/acids-ircam/variational-timbre}}.

\begin{figure}
\label{fig:Workflow}
\begin{center}
\includegraphics[scale=0.5]{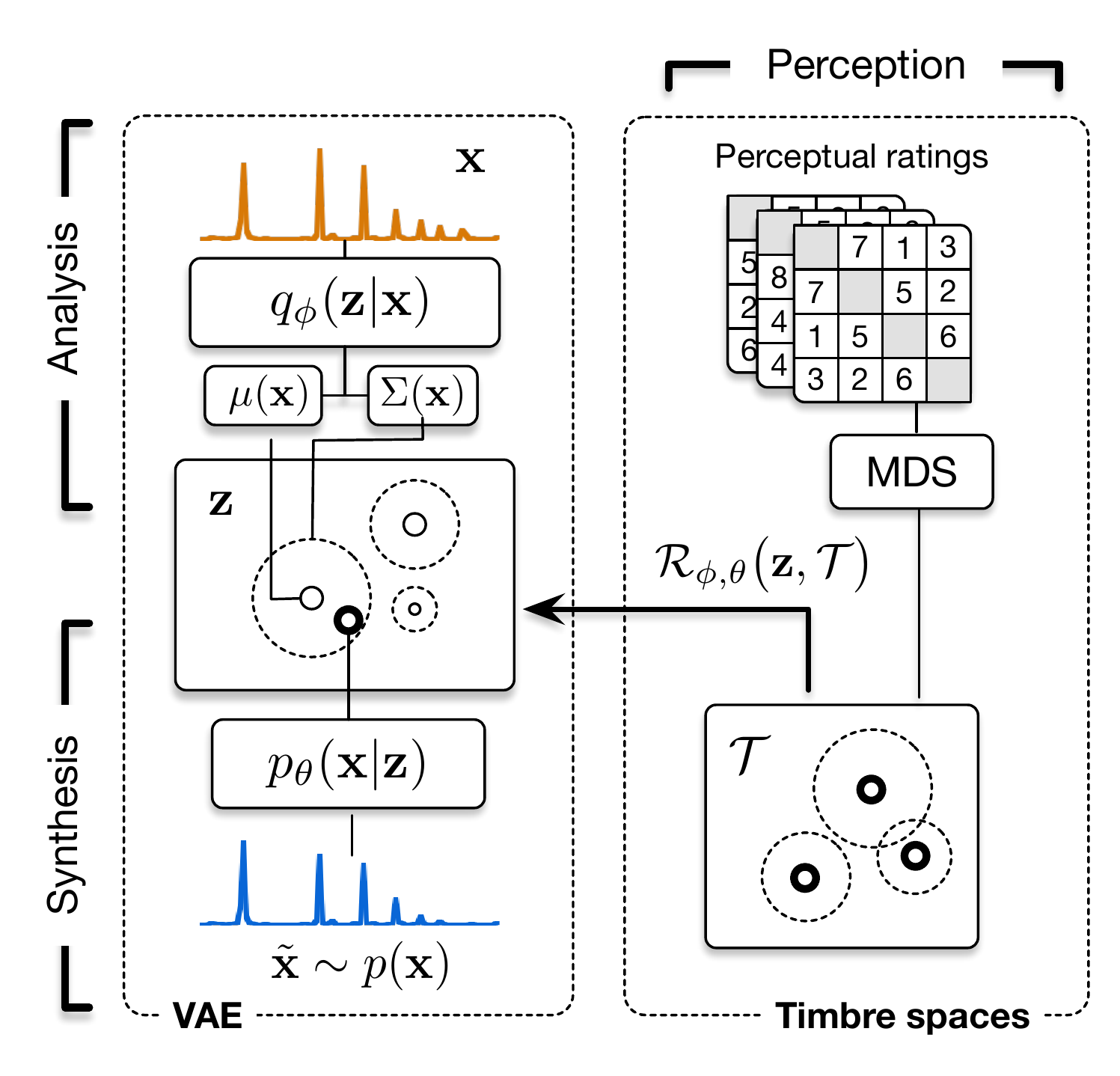}
\caption{(\emph{Left}) VAEs can model a spectral frame $\mathbf{x}$ of an audio sample by learning an encoder $q_\phi(\mathbf{z}\given\mathbf{x})$ which maps them to a Gaussian $\mathcal{N}(\mu(\mathbf{x})$, $\sigma(\mathbf{x}))$ inside a latent space $\mathbf{z}$. The decoder $p_\theta(\mathbf{x}\given\mathbf{z})$ samples from this Gaussian to generate a reconstruction $\tilde{x}$ of the spectral frame. (\emph{Right}) Perception studies use similarity ratings to construct \emph{timbre spaces} exhibiting perceptual distances between instruments. Here, we develop a regularization $\mathcal{R}(\mathbf{z}, \mathcal{T})$ enforcing that the variational model finds a topology of latent space $\mathbf{z}$ that matches the topology of the timbre space $\mathcal{T}$.}
\end{center}
\end{figure}

\section{State-of-art}

\subsection{Variational auto-encoders}

\textit{Generative models} are a flourishing class of learning approaches, which aim to find the underlying probability distribution of the data $p(\mathbf{x})$ \cite{bishop2014pattern}. Formally, based on a set of examples in a high-dimensional space $\mathbf{x}\in\mathbb{R}^{d_{x}}$, we assume that these follow an unknown distribution $p\left(\mathbf{x}\right)$. Furthermore, we consider a set of \emph{latent variables} defined in a lower-dimensional space $\mathbf{z}\in\mathbb{R}^{d_{z}}$ ($d_{z} \ll d_{x}$). These latent variables help govern the generation of the data and enhance the \emph{expressivity} of the model. Thus, the complete model is defined by the joint probability distribution $p(\mathbf{x}, \mathbf{z}) = p(\mathbf{x} \given \mathbf{z})p(\mathbf{z})$. We could find $p(\mathbf{x})$ through its relation to the posterior distribution $p(\mathbf{z} \given \mathbf{x})$ given by Bayes' theorem. However, for complex non-linear models (such as those that we will consider in this paper), this posterior can not be found in closed form.

For decades, the dominant paradigm for approximating $p(\mathbf{x})$ has been \emph{sampling} methods \cite{hastings1970monte}. However, the quality of this approximation depends on the number of sampling operations, which might be extremely large before we have an accurate estimate. Recently, \emph{variational inference} (VI) \cite{bishop2014pattern} has been proposed to solve this problem through \emph{optimization} rather than sampling. VI assumes that if the distribution is too complex to find, we could find a simpler approximate distribution that still models the data, while trying to minimize its difference to the real distribution. Formally, VI specifies a family $\mathcal{Q}$ of approximate densities, where each member $q(\mathbf{z}\given\mathbf{x})\in\mathcal{Q}$ is a candidate approximation to the exact $p\left(\mathbf{z} \given \mathbf{x}\right)$. Hence, the inference problem can be transformed into an optimization problem by minimizing the Kullback-Leibler (KL) divergence between the approximation and  original density
\begin{equation}
\label{eq:VariationalInference_KL}
q^{*}(\mathbf{z}\given \mathbf{x})=\argmin_{q(\mathbf{z} \given \mathbf{x})\in\mathcal{Q}} \mathcal{D}_{KL} \big[ q\left(\mathbf{z} \given \mathbf{x}\right) \parallel p\left(\mathbf{z} \given \mathbf{x}\right) \big]
\end{equation}
The complexity of the family $\mathcal{Q}$ will both determine the quality of the approximation, but also the complexity of this optimization. Hence, the major issue of VI is to choose $\mathcal{Q}$ to be flexible enough to closely approximate $p\left(\mathbf{z} \given \mathbf{x}\right)$, while being simple enough to allow efficient optimization. Now, if we expand the KL divergence that we need to minimize and rely on Bayes' rule to replace $p(\mathbf{z} \given \mathbf{x})$, we obtain the following expression
\begin{multline}
D_{KL} \big[ q(\mathbf{z} \given \mathbf{x}) \parallel p(\mathbf{z} \given \mathbf{x}) \big] = \mathbb{E}_{q(\mathbf{z})} \big[ \log{q(\mathbf{z} \given \mathbf{x})} - \log{p(\mathbf{x} \given \mathbf{z})} \\ 
 - \log{p(\mathbf{z})} + \log{p(\mathbf{x})}\big]
\end{multline}
Noting that the expectation is over ${q(\mathbf{z|x})}$ and that $p(\mathbf{x})$ does not depend on it, we can get this term out of the expectation and then observe that the remaining equation can be rewritten as another KL divergence leading to
\begin{multline}
\log{p(\mathbf{x})} - D_{KL} \big[ q(\mathbf{z} \given \mathbf{x}) \parallel p(\mathbf{z} \given \mathbf{x}) \big] = \\
\mathbb{E}_{\mathbf{z}} \big[ \log{p(\mathbf{x} \given \mathbf{z})}\big] - D_{KL} \big[ q(\mathbf{z} \given \mathbf{x}) \parallel p(\mathbf{z}) \big]
\end{multline}
This formulation describes the logarithm of the quantity that we want to maximize $\log p(\mathbf{x})$ minus the error we make by using an approximate $q$ instead of $p$. Therefore, we can optimize this alternative objective, called the \emph{evidence lower bound} (ELBO) as
\begin{equation}
\label{eq:VAE_Objective}
\log p(\mathbf{x}) = D_{KL} \big[q(\mathbf{z}\given\mathbf{x})\parallel p(\mathbf{z}\given\mathbf{x})\big] + ELBO(q).
\end{equation}
and the KL is non-negative, so $\log p(\mathbf{x}) \geq ELBO(q), \forall q(\mathbf{z})$. Now, to optimize this objective, we will rely on parametric distributions $q_{\phi}(\mathbf{z}\given \mathbf{x})$ and $p_{\theta}(\mathbf{x}\given\mathbf{z})$. Therefore, optimizing our generative model will amount to optimize these parameters $\big\{\theta,\phi\big\}$
\begin{equation}
\mathcal{L}(\boldsymbol{\theta, \phi}) = \mathbb{E}_{q_\phi (\mathbf{z})} \big[ \log{ p_\theta (\mathbf{x|z}) } \big] - D_{KL} \big[ q_\phi(\mathbf{z|x}) \parallel p_\theta(\mathbf{z}) \big]
\end{equation}
We can see that this equation involves $q_{\phi}(\mathbf{z} \given \mathbf{x})$ which \emph{encodes} the data $\mathbf{x}$ into the latent representation $\mathbf{z}$ and a \emph{decoder} $p(\mathbf{x} \given \mathbf{z})$, which generates a data $\mathbf{x}$ given a latent configuration $\mathbf{z}$. Hence, this whole structure defines the \emph{Variational Auto-Encoder} (VAE), which is depicted in Figure~1 (Left).

The VAE objective can be interpreted intuitively. The first term  increases the likelihood of the data generated given a configuration of the latent, which amounts to minimize the \emph{reconstruction error}. The second term represents the error made by using a simpler distribution $q_{\phi}(\mathbf{z} \given \mathbf{x})$ rather than the true distribution $p_{\theta}(\mathbf{z})$. Therefore, this \emph{regularizes} the choice of approximation $q$ so that
\begin{equation}
\mathcal{L}_{\boldsymbol{\theta, \phi}} = \underbrace{\mathbb{E}_{q_\phi (\mathbf{z})} \big[ \log{ p_\theta (\mathbf{x|z}) } \big]}_{\text{reconstruction}} - \beta \cdot\underbrace{D_{KL} \big[ q_\phi(\mathbf{z|x}) \parallel p_\theta(\mathbf{z}) \big]}_{\text{regularization}} 
\end{equation}
The first term can be optimized through a usual maximum likelihood estimation, while the second term requires that we define the prior $p(\mathbf{z})$. While the easiest choice is to choose $p(\mathbf{z})\sim\mathcal{N}(\mathbf{0},\mathbf{I})$, it also adds the benefit that this term has a simple closed solution for computing the optimization, as detailed in \cite{kingma2013auto}. Here we introduced a weight $\beta$ to the KL divergence, which leads to the $\beta$-VAE formulation \cite{higgins2016beta}. This has been shown to improve the capacity of the model to disentangle factors of variations in the data. However, it has later been shown that an appropriate way to handle this parameter was to perform \emph{warm-up} \cite{sonderby2016train}, where the $\beta$ parameter is linearly increased in the first epochs of training.

Finally, we need to select a family of variational densities $\mathcal{Q}$. One of the most widespread choice is the \emph{mean-field variational family} where latent variables are independent and are each parametrized by a distinct variational parameter
\begin{equation}
q(\mathbf{z})=\prod_{j=1}^{m} q_{j}(z_{j})
\end{equation}
Therefore, each dimension of the latent space will be governed by an independent Gaussian distribution with its own mean and variance depending on the input data $q_j(z_j) = \mathcal{N}(\mu_{j}(\mathbf{x}),\Sigma_{j}(\mathbf{x}))$.

VAEs are powerful representation learning frameworks, while remaining simple and fast to learn without requiring large sets of examples \cite{sonderby2016train}. Their potential for audio applications have been only scarcely investigated yet and mostly in topics related to speech processing such as blind source separation \cite{kuo2017variational} and speech transformation \cite{hsu2017learning}. However, to the best of our knowledge, the use of VAE and their latent spaces to perform musical audio analysis and generation has yet to be investigated.

\subsection{Timbre spaces and auditory perception}
\label{sec:SOA-Timbre-spaces}
For several decades, music perception research has tried to understand the mechanisms leading to the perception of \emph{timbre}. Several studies have shown that timbre could be partially described by computing various audio descriptors \cite{mcadams1995perceptual}. To do so, most studies relied on the concept of \emph{timbre spaces} \cite{grey1978perceptual}, a model that organize audio samples based on perceptual dissimilarity ratings. In these studies, pairs of sounds are presented to subjects that are asked to rate their perceptual dissimilarities inside a given set of instruments. Then, these ratings are compiled into a set of dissimilarity matrices that are analyzed with Multi-Dimensional Scaling (MDS). The MDS algorithm provides a timbre space that exhibits the underlying perceptual distances between different instruments (Figure~1 (Right)). Here, we briefly detail corresponding studies and redirect interested readers to the full articles for more details.\\
In his seminal paper, Grey \cite{grey1977multidimensional} performed a study with 16 instrumental sound samples. Each of the 22 subjects had to rate the dissimilarity between all pairs of sounds on a continuous scale from 0 (most similar) to 1 (most dissimilar). This lead to the first construction of a timbre space for instrumental sounds. They further exhibit that the dimensions explaining these dissimilarities could be correlated to the spectral centroid, spectral flux and attack centroid. Several studies followed this research by using the same experimental paradigm. Krumhansl \cite{krumhansl1989musical} used 21 instruments with 9 subjects on a discrete scale from 1 to 9, Iverson et al. \cite{iverson1993isolating} with 16 samples and 10 subjects on a continuous scale from 0 to 1, McAdams et al. \cite{mcadams1995perceptual} with 18 orchestral instruments and 24 subjects on a discrete scale from 1 to 16 and, finally, Lakatos \cite{lakatos2000common} with 17 subjects on 22 harmonic and percussive samples on a continuous scale from 0 to 1.
Each of these studies shed light on different aspects of audio perception, depending on the aspect being scrutinized and the interpretation of the space by the experimenters. However, all studies have led to different spaces with different dimensions. The fact that different studies correlate to different audio descriptors prevents a generalization of the acoustic cues that might correspond to timbre dimensions. Furthermore, timbre spaces have been explored based on MDS to organize perceptual ratings and correlate spectral descriptors \cite{mcadams1995perceptual}. Therefore, these studies are inherently limited by the fact that
\begin{itemize}
\item ordination techniques (such as MDS) produce fixed spaces that must be recomputed for any new data point
\item these spaces do not generalize nor synthesize audio between instruments as they do not provide an invertible mapping
\item interpretation is bounded to the \emph{a posteriori} linear correlation of audio descriptors to the dimensions rather than analyzing the topology of the space itself
\end{itemize}

As noted by McAdams et al. \cite{mcadams2006meta}, critical problems in these approaches are the lack of an objective distance model based on perception and general dimensions for the interpretation of timbral transformation and source identification. Here, we show that relying on VAE models to learn unsupervised spaces, while regularizing the topology of these spaces to fit given perceptual ratings can allow to alleviate all of these limitations.

\section{Regularizing latent space topology}

In this paper, we aim to construct a latent space that could both analyze and synthesize audio content, while providing the underlying perceptual relationships between audio samples. To do so, we show that we can influence the organization of the VAE latent space $\mathbf{z}$ so that it follows the topology of a given target space $\mathcal{T}$. Here, we will rely on the MDS space constructed from perceptual ratings as a target space $\mathcal{T}$. However, it should be noted that this idea can be applied to any given target space that provides a set of distances between the elements used for learning the VAE space.

To further specify our problem, we consider a set of audio samples, where each $x_{i}$ can be encoded in the latent space as $\mathbf{z}_{i}$ and have an equivalent in the target space $\mathcal{T}_{i}$. In order to relate the elements of the audio dataset to the perceptual space, we consider that each sample is labeled with its instrumental class $\mathcal{C}_{i}$, that has an equivalent in the timbre space. Therefore, we will match the properties of the classes between the latent and target spaces (note that we could use element-wise properties for finer control). 

Here, we propose to regularize the learning by introducing the perceptual similarities through an additive term $\mathcal{R}\left(\mathbf{z},\mathcal{T}\right)$. This \emph{penalty} imposes that the properties of the latent space $\mathbf{z}$ are similar to that of the target space $\mathcal{T}$. The optimization objective becomes
\begin{equation}
\mathbb{E}\big[ \log{ p_{\theta}(\mathbf{x|z}) } \big] - \beta D_{KL} \big[ q_{\phi}(\mathbf{z|x}) \parallel p_{\theta}(\mathbf{z}) \big]+\alpha \mathcal{R}\big(\mathbf{z},\mathcal{T}\big)
\end{equation}
where $\alpha$ is an hyper-parameter that allows us to control the influence of the regularization. Hence, amongst two otherwise equal solutions, the model is pushed to select the one that comply with the penalty. In our case, we want the distances between instruments to follow perceptual timbre distances. Therefore, we need to minimize the differences between the set of distances in the latent space $\mathcal{D}_{i,j}^\mathbf{z}=\mathcal{D}(\mathbf{z}_{i}, \mathbf{z}_{j})$ and the distances in target space $\mathcal{D}_{i,j}^\mathcal{T}=\mathcal{D}(\mathcal{T}_{i}, \mathcal{T}_{j})$. Therefore, the regularization criterion will try to minimize the overall differences between these sets of distances.
To compute these sets, we take inspiration from the \emph{t-Stochastic Neighbor Embedding} (t-SNE) algorithm \cite{maaten2008visualizing}. Indeed, as their goal is to map the distances from one (high-dimensional) space into a target (low-dimensional) space, it is highly correlated to our task. However, we can not simply apply the t-SNE algorithm on the latent space as this would lead to a non-invertible mapping. Instead, we aim to steer the learning in a similar way. Hence, we compute the relationships in the latent space $\mathbf{z}$ by using the conditional Gaussian density that $i$ would choose $j$ as its neighbor
\begin{equation}
\mathcal{D}^\mathbf{z}_{i,j}=\frac{\exp\big(-\norm{\mathbf{z}_{i} - \mathbf{z}_{j}}^2/2\sigma^2_{i}\big)}{\sum_{k\neq i}\exp\big(-\norm{\mathbf{z}_{i} - \mathbf{z}_{k}}^2/2\sigma^2_{i}\big)}
\end{equation}
where $\sigma_i$ is the variance of the Gaussian centered on $\mathbf{z}_i$, defined as $\sigma_i=1/\sqrt{2}$. Then, to relate the points in the timbre space $\mathcal{T}$, we use a Student-t distribution to define the distances in this space as
 \begin{equation}
\mathcal{D}^\mathcal{T}_{i,j} = \frac{\big(1+\norm{\mathcal{T}_{i} - \mathcal{T}_{j}}^2\big)^{-1}}{\sum_{k\neq l}\big(1+\norm{\mathcal{T}_{k} - \mathcal{T}_{l}}^2\big)^{-1}}
\end{equation}
Finally, we rely on the sum of KL divergences between the two distributions of distances in different spaces to define our complete regularization criterion
$$
\mathcal{R}\big(\mathbf{z}, \mathcal{T}\big)=\sum_{i}\mathcal{D}_{KL}\big[\mathcal{D}_{i}^{\mathbf{z}} \parallel \mathcal{D}_{i}^{\mathcal{T}}\big] = \sum_{i}\sum_{j}\mathcal{D}_{i,j}^{\mathbf{z}}\log\frac{\mathcal{D}_{i,j}^{\mathbf{z}}}{\mathcal{D}_{i,j}^{\mathcal{T}}}
$$

Hence, instead of applying a distance minimization a posteriori, we steer the learning to find a configuration of the latent space $\mathbf{z}$ that displays the same distance properties as the space $\mathcal{T}$, while providing an invertible mapping.

\section{Experiments}

\subsection{Datasets}
\label{sec:datasets}
\begin{figure}
\begin{center}
\includegraphics[scale=0.35]{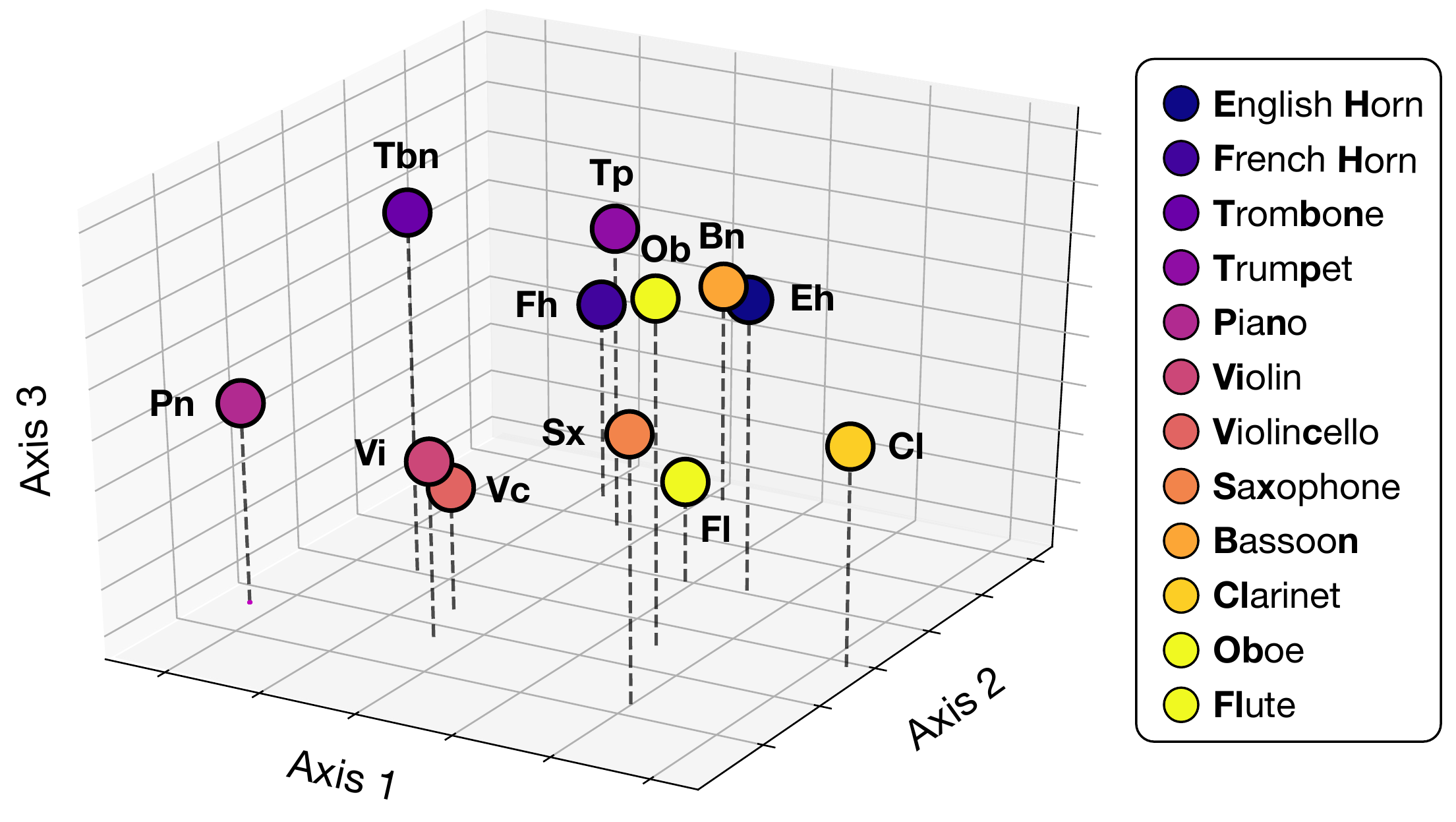}
\end{center}
\caption{Multi-dimensional scaling (MDS) of the combined and normalized set of perceptual ratings from different studies.}
\label{fig:MDS-Space}
\end{figure}

\emph{Timbre studies.}
We rely on the perceptual ratings collected across five independent timbre studies \cite{grey1977multidimensional, krumhansl1989musical, iverson1993isolating, mcadams1995perceptual, lakatos2000common}. As discussed earlier, even though all studies follow the same experimental protocol, there are some discrepancies in the choice of instruments, rating scales and sound stimuli. However, here we aim to obtain a consistent set of properties to define a common timbre space. 
Therefore, we computed the maximal set of instruments for which we had ratings for all pairs. To do so, we collated the list of instruments from all studies and counted their co-occurences, leading to a set of 12 instruments (Piano, Cello, Violin, Flute, Clarinet, Trombone, French Horn, English Horn, Oboe, Saxophone, Trumpet, Tuba) with pairwise ratings.
Then, we normalized the raw dissimilarity data (keeping all instruments of that study) so that it maps to a common scale from 0 to 1. Finally, we extracted the set of ratings that corresponds to our selected instruments. This leads to a total of 1217 subject ratings for all instruments, amounting to 11845 pairwise ratings. Based on this set of ratings, we compute an MDS space to ensure the consistency of our normalized perceptual space on the selected set. The results of this analysis are displayed in Figure~\ref{fig:MDS-Space}. We can see that even though the ratings come from different studies, the resulting space remains very coherent, with the distances between instruments remaining coherent with the original perceptual studies.

\emph{Audio datasets.}
In order to learn the distribution of instrumental sounds directly from the audio signal, we rely on the Studio On Line (SOL) database \cite{ballet1999studio}. We selected 2,200 samples to represent the 11 instruments for which we extracted perceptual ratings. We normalized the range of notes used by taking the whole tessitura and dynamics available (to remove effects from the pitch and loudness). All recordings were resampled to 22050 Hz for the experiments. 
Then, as we intend to evaluate the effect of different spectral distributions as input to our proposed model, we computed several invertible transforms for each audio sample. First, we compute the Short-Term Fourier Transform (STFT) with a Hamming window of 40ms and a hop size of 10ms. Then, we compute the Discrete Cosine Transform (DCT) with the same set of parameters. Finally, we compute the Non-Stationary Gabor Transform (NSGT) \cite{balazs2011theory} mapped either on a Constant-Q scale of 48 bins per octave and a Mel scale or ERB scale of 400 bins, all from 30 to 11000 Hz. For all transforms, we only keep the magnitude of the distribution to train our models. We perform a corpus-wide normalization to preserve the relative intensities of the samples (normalizing all distributions by the maximal value found across samples). Then, we extract a single temporal frame from the sustained part of the representation (200 ms after the beginning of the sample) to represent a given audio sample. Finally, the dataset is randomly split across notes to obtain a training (90\%) and test (10\%) set.

\emph{Audio reconstruction.} To perform audio synthesis, we consider paths inside the latent space, where each point corresponds to a single spectral frame. We sample along a given path and concatenate the spectral frames to obtain the magnitude distribution. Then, we apply the Griffin-Lim algorithm in order to recover the phase distribution and synthesize the corresponding waveform.

\subsection{Models}
Here, we rely on a simple VAE architecture to show the efficiency of the proposed method. The encoder is defined as a 3-layer feed-forward neural network with Rectified Linear Units (ReLU) activation functions and 2000 units per layer. The last layer maps to a given dimensionality $d$ of the latent space. In our experiments, we analyzed the effect of relying on different latent spaces and empirically selected latent spaces with 64 dimensions. The decoder is defined in a symmetrical way, with the same architecture and units, mapping back to the dimensionality of the input transform. For learning the model, we use a value of $\beta=2$, which is linearly increased from 0 to its final value during the first 100 epochs (following the \emph{warmup} procedure \cite{sonderby2016train}). In order to train the model, we rely on the ADAM \cite{kingma2014adam} optimizer with an initial learning rate of 0.0001. In a first stage, we train the model without perceptual regularization ($\alpha=0$) for a total of 5000 epochs. Then, we introduce the perceptual regularization ($\alpha=0.1$) and train for another 1000 epochs. This allows the model to first focus on the quality of the reconstruction, and then to converge towards a solution with perceptual space properties. We found in our experiments that this two-step procedure is critical to the success of the regularization.

\section{Results}


\subsection{Latent spaces properties}
\label{sec:Results_spaces}

\begin{figure*}
\begin{center}
\includegraphics[scale=0.28]{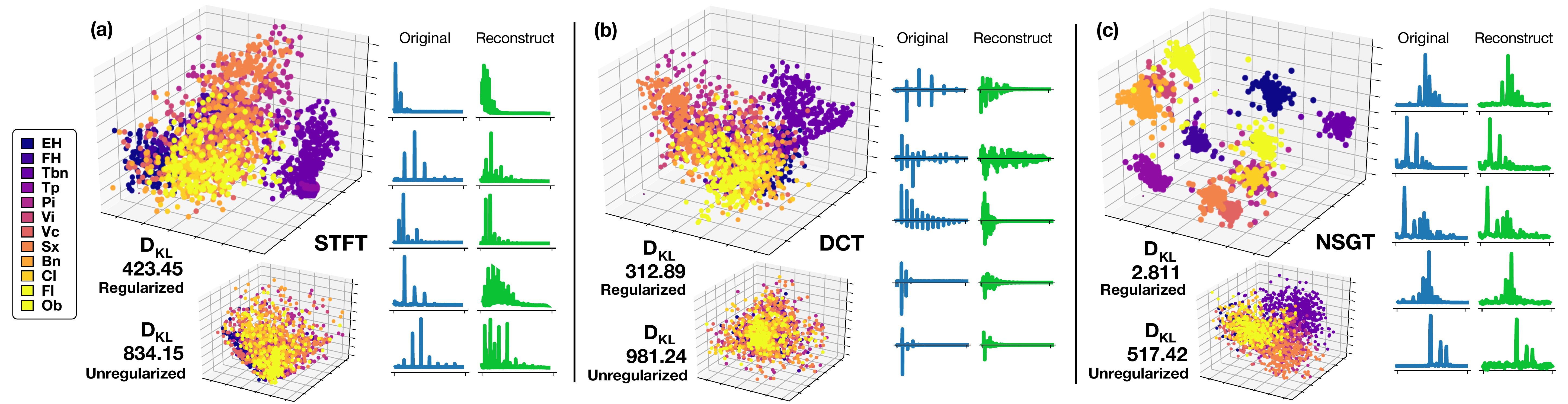}
\end{center}
\caption{Comparing the regularized VAE latent spaces (large) for the STFT (a), DCT (b) and NSGT (CQT) (c) transforms. For each transform, we plot the corresponding unregularized space (small) and their respective $\mathcal{D}_{KL}$ divergence to the timbre space distances. We plot a set of VAE decoder reconstructions of instrumental spectral frame distributions from the test set directly from the regularized spaces}
\label{fig:Regularized_spaces}
\end{figure*}

In order to visualize the 64d latent spaces, we apply a simple Principal Component Analysis (PCA) to obtain a 3d representation. Using a PCA ensures that the visualization is a linear transform of the original space. Therefore, this preserves the real distances inside the latent space. Furthermore, this will allow to recover an exploitable representation when we will use this space to generate novel audio content. The results of learning regularized latent spaces for different spectral transforms are displayed in Figure~\ref{fig:Regularized_spaces}.

As we can see, in VAEs without regularization (small space), the relationships between instruments do not match perceptual ratings. Furthermore, the variance of distributions show that the model rather tries to spread the information across the latent space to help the reconstruction. However, the NSGT provides a better unregularized space with different instrumental distributions already well separated.
Now, if we compare to the regularized spaces, we can clearly see the effect of the criterion, which provides a larger separation of distribution. This effect and final result is particularly striking for the NSGT (c), which provides the highest correlation to the distances in our combined timbre space (Figure~\ref{fig:MDS-Space}). Interestingly, the instrumental distributions might be shuffled around space in order to comply with the reconstruction objective. However, the pairwise distances reflecting perceptual relations are well matched as indicated by the KL divergence. By looking at the test set reconstructions, we can see that enforcing the perceptual topology on the latent spaces do not impact the quality of audio reconstruction for the NSGT, where the reconstruction provides an almost perfectly matching distribution. In the case of the STFT, we can see that the model is impacted by the regularization and mostly match the overall density of the distribution rather than its exact peak information. Finally, it seems that the DCT model diverged in terms of reconstruction, being unable to reconstruct the distributions. However, we can see that the KL fit to timbre distances is better than the STFT, indicating an overfit of the learning towards the regularization criterion. This generative evaluation is quantified and confirmed in the next section.

\subsection{Generative capabilities}
\label{sec:Results_Generative}
We quantify the generative capabilities from the latent spaces by computing the log likelihood and mean difference between the original and reconstructed spectral representations on the test set. We compare these results for different transforms and without regularization, which are presented in Table~\ref{tab:Generative}.

\begin{table}
\small
\begin{center}
\begin{tabular}{c|c|c|c}
\hline 
\multicolumn{2}{c|}{Method} & $\log p(\mathbf{x})$ & $\norm{\mathbf{x}-\tilde{\mathbf{x}}}^2$ \tabularnewline
\hline 
\hline
\multirow{3}{*}{\makecell{Unregularized \\ (NSGT)}} & PCA & - & 2.2570 \tabularnewline
& AE & -1.2008 & 1.6223 \tabularnewline
& VAE & -2.3443 & 0.1593\tabularnewline
\hline 
\multirow{5}{*}{\makecell{Regularized \\ (VAE)}} & STFT & -1.9237 & 0.2412\tabularnewline
& DCT & 4.3415 & 2.2629\tabularnewline
& NSGT-CQT & -2.8723 & 0.1610\tabularnewline
& NSGT-MEL & -2.9184 & 0.1602\tabularnewline
& NSGT-ERB & \textbf{-2.9212} & \textbf{0.1511}\tabularnewline
\hline 
\end{tabular}
\end{center}
\caption{Generative capabilities evaluated by the log likelihood and mean quality of reconstructed representations on the test set.}
\label{tab:Generative}
\end{table}
As we can see, the unregularized VAE trained on the NSGT distribution provides a very good reconstruction capacity, and still generalizes very well. This can be seen in its ability to generate spectral distributions from the test set almost perfectly. Interestingly, regularizing the latent space does not seem to affect the quality of the reconstruction at all. It even seems that the generalization increases with the regularized latent space. This could however be explained by the fact that the regularized models are trained for twice as much epochs based on our two-fold procedure.

It clearly seems that NSGTs provide both better generalization and reconstruction abilities, while the DCT seems to provide only a divergent model. This can be explained by the fact that NSGT frequency axis is organized on a logarithmic scale. Furthermore, their distribution are well spread across this axis, whereas STFT and DCT tends to have most of their informative dimensions in the bottom half of the spectrum. Therefore, NSGTs provide a more informative input. Finally, there only seems to be a marginal difference between the results of different NSGT scales. However, for all remaining experiments, we select the NSGT-ERB as it is more coherent with our perceptual endeavor.

Thanks to the decoder and its generative capabilities, we can now directly synthesize the audio corresponding to any point inside the latent space, but also any paths between two given instruments. This allows us to turn our analytical spaces into audio synthesizers. Furthermore, as shown in Figure~\ref{fig:Descriptors_Topology} (Bottom right), synthesizing audio along these spaces lead to smooth evolution of spectral distributions and perceptually continuous synthesis (as discussed extensively in the next section). In order to perform subjective evaluation of the audio reconstruction, generated samples from the latent space are available on the supporting repository.

\begin{figure}
\begin{center}
\includegraphics[scale=0.35]{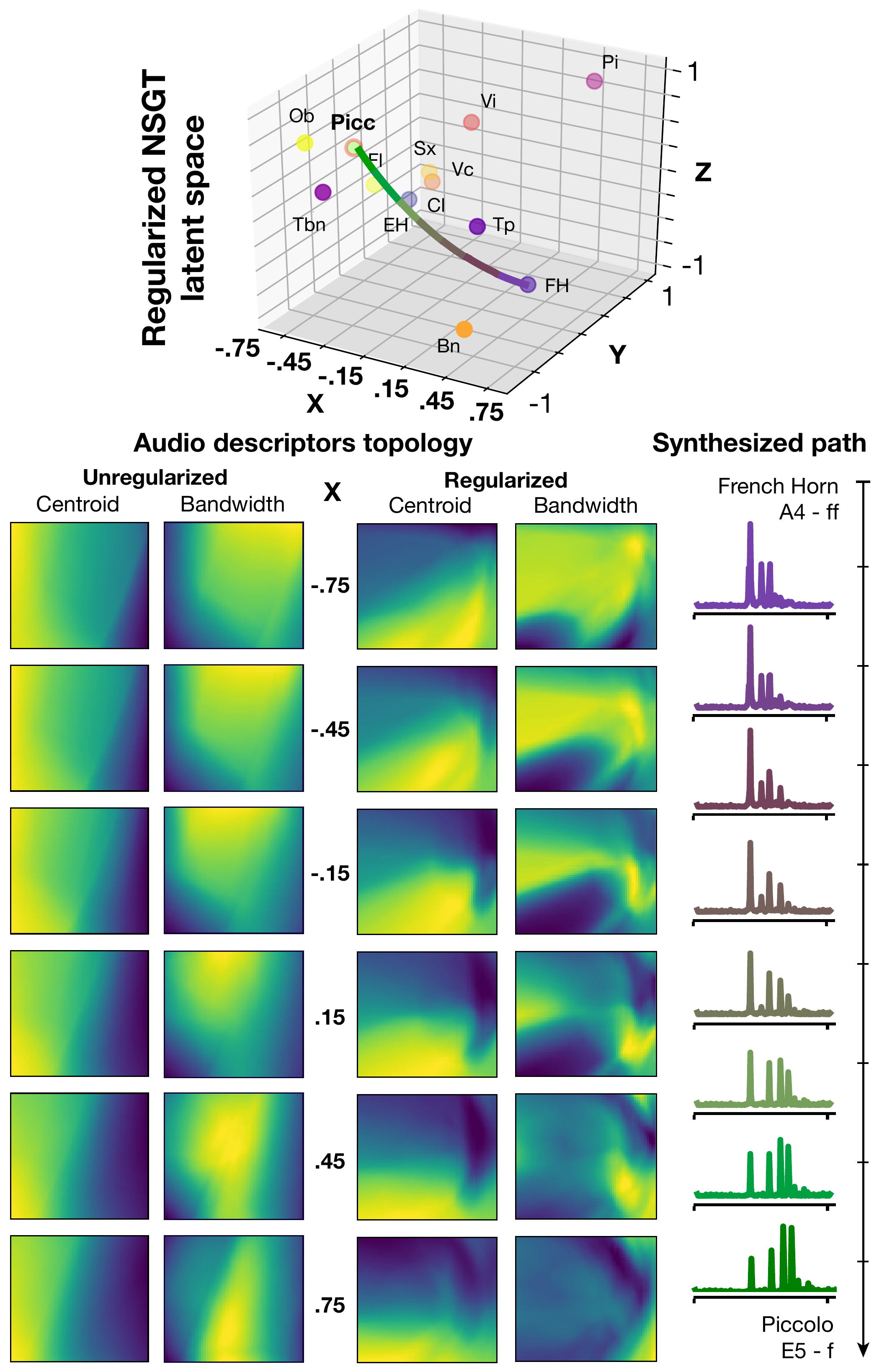}
\end{center}
\caption{(Top) Projecting new instruments inside the regularized latent space allow to see their perceptual relations to others. (Bottom right) We can generate any path between instruments in the space and synthesize the corresponding perceptually-smooth audio evolution. (Bottom, left)  We define 6 equally-spaced projection planes across the $x$ axis and sample points on a 50x50 grid. We reconstruct their audio distribution to compute their spectral \emph{centroid} and \emph{bandwidth}. We compare the resulting descriptor space topology for unregularized (left) and regularized (right) spaces.}
\label{fig:Descriptors_Topology}
\end{figure}

\subsection{Generalizing perception, audio synthesis of timbre paths}
\label{sec:Results_Perceptual}
Given that the encoder of our latent space is trained directly on spectral distributions, it is able to analyze samples belonging to new instruments that were not part of the original perceptual studies. Furthermore, as the learning is regularized by perceptual ratings, we could hope that the resulting position would predict the perceptual relationships of this new instrument to the existing instruments. This could potentially feed further perceptual studies, to refine timbre understanding. To evaluate this hypothesis, we extracted a set of \emph{Piccolo} audio samples to evaluate their behavior in latent space. We perform the same processing as for the training dataset (Section~\ref{sec:datasets}) and encode these new samples in the latent space to study the \emph{out-of-domain} generalization capabilities of our model. The results of this analysis are presented in Figure~\ref{fig:Descriptors_Topology} (Top).

Here, we can see that new samples (represented by their centroid for clarity) are encoded in a coherent position in the latent space, as they group with their families, even though they were never presented to the model during learning. However, obtaining a definitive answer on the perceptual inference capabilities of these spaces would require a complete perception experiment, that we leave to future work.
Now, as argued previously, one of the key property of the latent spaces is that they provide an invertible non-linear mapping. Therefore, we could thrive on this property to truly understand what are the perceptual relations between instruments based on the behavior of spectral distributions \emph{between} the points in the timbre space. To exhibit this capability, we encode the position in the latent space of a \emph{Piccolo} sample playing an E5-f. Then, based on the position of a \emph{French Horn} playing an A4-ff, we perform an interpolation between these latent points to obtain the path between these two instruments in latent space. We then sample and decode the spectral distributions at 6 equally spaced positions along the path, which are displayed in Figure~\ref{fig:Descriptors_Topology} (Right). As we can see, the resulting audio distributions demonstrate a smooth evolution between timbral structures of both instruments. Furthermore, the resulting interpolation is clearly more complex than a linear change between one structure to the other. Hence, this approach could be used to understand more deeply the timbre relationships between instruments. Also, this provides a model able to perform perceptually-relevant synthesis of novel timbres, while sharing the properties of multiple instruments.

\subsection{Topology of audio descriptors}
\label{sec:Results_Topology}
Here, we analyze the topology of signal descriptors across the latent space. As the space is continuous, we do so by sampling uniformly the PCA space and then using the decoder to generate audio samples at a given point. Then, we compute the audio descriptors of this sample. In order to provide a visualization, we select 6 equally-distant planes across the $x$ dimension, at $\{-.75, -.45, -.15, .15, .45, .75\}$, which define an uniform 50x50 grid between $[-1, 1]$ on other dimensions. We compare the results between unregularized or regularized NSGT latent spaces in Figure~\ref{fig:Descriptors_Topology} (Bottom left) for the \emph{spectral centroid} and \emph{spectral bandwidth}. Animations of continuous traversals of the latent space are available on the supporting repository.
As we can see, the audio descriptors behave following overall non-linear patterns for both unregularized and regularized latent spaces. However, they still exhibit locally smooth properties. This shows that our model is able to organize audio variations. In the case of unregularized spaces, the organization of descriptors is spread out in a more even fashion. The addition of perceptual ratings to regularize the learning seems to require that this space is organized with a more complex topology. This could be explained by the fact that, in the unregularized case, the VAE only needs to find a configuration of the distributions that maximizes their reconstruction. Oppositely, the regularization requires that instrumental distances follow the perceptual dissimilarity ratings, prompting the need for a more complex relationship between descriptors. This might underline the fact that linear correlations between MDS dimensions and audio descriptors is insufficient to truly understand the dimensions related to timbre perception. However, the audio descriptors topology overall still provide locally smooth evolutions. Finally, a very interesting observation comes from the topology of the centroid. Indeed, all perceptual studies underline its correlation to timbre perception, which is partly confirmed by our model (by projecting on the y axis). This tends to confirm the perceptual relevance of our regularized latent spaces. However, this also shows that the relation between centroid and timbre might not be linear.

\subsection{Descriptor-based synthesis}
\label{sec:Results_Descriptor_Synthesis}

As shown in the previous section, the audio descriptors are organized in a smooth locally linear way across the space. Furthermore, as discussed in Section~\ref{sec:Results_spaces}, we have seen that the instrumental distributions are grouped across spaces depending on perceptual relations. Based on these two findings, we hypothesize that we can find paths inside these spaces that modify a given audio distribution to follow a target descriptor, while remaining perceptually smooth. Hence, we propose a simple method for perceptually-relevant \emph{descriptor-based path synthesis} presented in Algorithm~\ref{algo:Desc_synth}.

\begin{algorithm}
\SetAlgoLined
\KwData{space $\mathbf{z}$, encoder $q_\phi(\mathbf{z|x})$, decoder $p_\theta(\mathbf{x|z})$}
\KwData{origin spectrum $\mathbf{x}_{0}$, target series $\mathbf{t}_{1..N}$, descriptor $d$}
 \KwResult{spectral distrib. $S\in\mathbb{R}^{N\times F}$}
 // Find origin position in latent space\\
 $\mathbf{z}_{0} = q_\phi(\mathbf{x}_{0})$\\
 // Evaluate origin descriptor\\
 $\mathbf{d}_{0} = evaluate(\mathbf{x}_{0}, d)$\\

 \For{$i\in[1,N]$}{
  // Latent 3-d neighborhood of current point\\
  $\mathbf{N}_{i} = neighborhood(\mathbf{z}_{i-1})$\\
  // Sample and evaluate descriptors\\
  $\mathbf{X}_{i} = q_\phi(\mathbf{N}_{i})$\\
  $\mathbf{D}_{i} = evaluate(\mathbf{X}_{i}, d)$\\
  // Compute difference to target\\
  $\Delta_{i} = \norm{(\mathbf{D}_i - \mathbf{d}_{i-1}) - (t[i]-t[i-1])}^{2}$\\
  // Find next latent point \\
  $\mathbf{z}_{i} = argmin(\Delta_{i})$\\
  // Decode distribution\\
  $S[i] = p_\theta(\mathbf{z}_{i})$
 }

\caption{Descriptor-based path synthesis}
\label{algo:Desc_synth}
\end{algorithm}

Based on the latent space $\mathbf{z}$ (with corresponding encoder $q$ and decoder $p$) and a given origin spectrum $\mathbf{x}_0$, the goal of this algorithm is to find the succession of spectral distributions that match a given target evolution $\mathbf{t}\in\mathbb{R}^N$ for a descriptor $d$. First, we find the position of the origin distribution in latent space $\mathbf{z}_{0}$ and evaluate its descriptor value $\mathbf{d}_{0}$ (lines 1-4). Then for each point $i$, we compute the descriptor values $\mathbf{D}_i$ in the neighborhood of the current latent point (lines 6-10) by decoding their audio distributions. Note that the the neighborhood is defined as the set of close latent points, and its size directly defines the complexity of the optimization. Then, we select the neighboring latent point $\mathbf{z}_i$ that provides the evolution of descriptor closest to the target evolution $t[i]$ (lines 11-14). Finally, we obtain the spectral distribution $S[i]$ by decoding the latent position $\mathbf{z}_i$. The results of applying this algorithm to a given instrumental distribution is presented in Figure~\ref{fig:Descriptors_Topology}.

\begin{figure*}
\begin{center}
\includegraphics[scale=0.4]{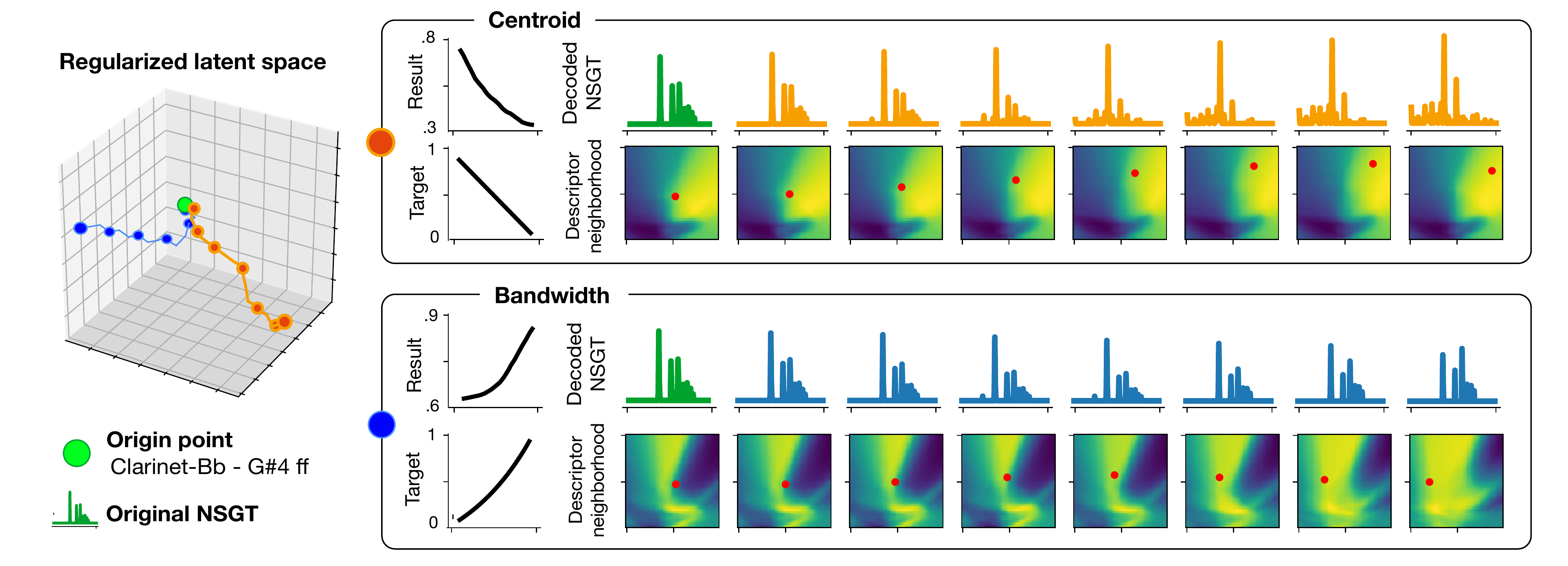}
\end{center}
\caption{\emph{Descriptor-based synthesis}. Given an origin point in latent space (Clarinet-Bb G\#4 ff), we apply our algorithm either on a descending target shape for the spectral centroid (top), or an ascending log shape for the spectral bandwidth (bottom). In both cases, we plot the decoded NSGT distributions and neighboring descriptor space information along the optimized path}
\label{fig:Descriptors_Topology}
\end{figure*}

Here, we start from the NSGT distribution of a Clarinet-Bb playing a G\#4 in \emph{fortissimo}. We apply our algorithm twice from the same origin point, either on a descending target shape for the spectral centroid (top), or an ascending log shape for the spectral bandwidth (bottom). In both cases, we plot the synthesized NSGT distributions at different points of the optimized path, and the neighboring descriptor space. As we can see, the resulting descriptor evolution closely match the input target in both cases. Furthermore, we can see by visual inspection of the spectrum evolution, that the corresponding distributions are indeed sharply modified to match the desired descriptors. Interestingly, the optimization of different target shapes on different descriptors lead to widely different paths in the latent space. However, the overall timbre structure of the original instrument still seems to follow a smooth evolution. Here, we note that the algorithm is quite rudimentary, and could benefit from more global neighborhood information, as witnessed from the slightly erratic local selection of latent points.

\section{Conclusion}

Here, we have shown that regularizing VAEs with perceptual ratings provides timbre spaces that allow for high-level analysis and audio synthesis directly from these spaces. The organization of these perceptually-regularized latent spaces prove the flexibility of these systems, and provides a latent space from which generation of novel audio content is straightforward. These spaces allow to extrapolate perceptual results on new sounds and instruments without the need to collect new measurements. Finally, by analyzing the behavior of audio descriptors across the latent space, we have shown that even though they follow a non-linear evolution, they still exhibit some locally smooth properties. Based on these, we introduced a method for descriptor-based path synthesis that allow to synthesize audio that match a target descriptor shape, while retaining the timbre structure of instruments. Future work on these latent spaces would be to perform perceptual experiments to confirm their perceptual topology.

\bibliographystyle{IEEEbib}
\bibliography{main}

\end{document}